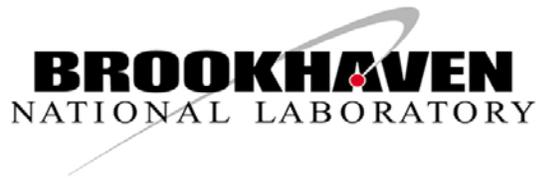

# Double Quarter Wave Crab Cavity Wire Stretching Measurement at BNL

Q. Wu

September 2020

Electron-Ion Collider

**Brookhaven National Laboratory**

**U.S. Department of Energy**

USDOE Office of Science (SC), Nuclear Physics (NP) (SC-26)

# DISCLAIMER

# *Double Quarter Wave Crab Cavity Wire Stretching Measurement at BNL*

Date: September 18th, 2020

Author: Qiong Wu, Tianmu Xin, Binping Xiao

Location: Brookhaven National Laboratory Building 912

## Introduction:

The wire stretching measurement was completed on the prototype Double Quarter Wave (DQW) crab cavity for operation practice and calibration of the measurement system. Four locations were defined to be on the electrical center plane of the crab cavity, and survey of the wire indicated all are on the same plane. The successful measurement validated the wire stretching system built at Brookhaven National Lab. The offset of the four wire locations to the fitted plane provided the error of the measurement.

## Background:

The Double Quarter Wave (DQW) crab cavity is one out of the two crab cavities within Work Package 4 of the LHC (Large Hadron Collider) Hi-Lumi Program, and the RF design of the cavity was originally supported by the US LHC Accelerator Research Program (LARP) and later the LHC Accelerator Upgrade Program (AUP) funded by the DOE High Energy Physics Office in the Office of Science. Brookhaven National Laboratory (BNL) has delivered the successful DQW crab cavity RF design, and worked actively with all collaborators and manufacturers afterwards with the fabrication and testing of the prototype cavities.

The wire stretching measurement was developed at Jefferson Lab, and the mechanism details of the system can be found in Ref. [1]. The main purpose of the wire stretching measurement is to find out the electrical center of the RF cavity. This calibration allows the cavity electrical center to be aligned with the beam trajectory during operation, which eliminated the fabrication error and asymmetrical factor between the cavity electrical center and geometrical center.

The stretch wire system used for this measurement was assembled at BNL, and tested with other copper cavities previously [2].

## Measurement:

The DQW cavity was supported and constrained on an optical table with fixtures and bolts. A thin brass wire coated with Zinc with a diameter of 0.25 mm was pulled through the cavity beam pipe and stretched with maximum tension in the assembly. One end of the wire was connected to the network analyzer which sent an RF signal down the wire and through the cavity. The RF signal is generated and controlled by a network analyzer.

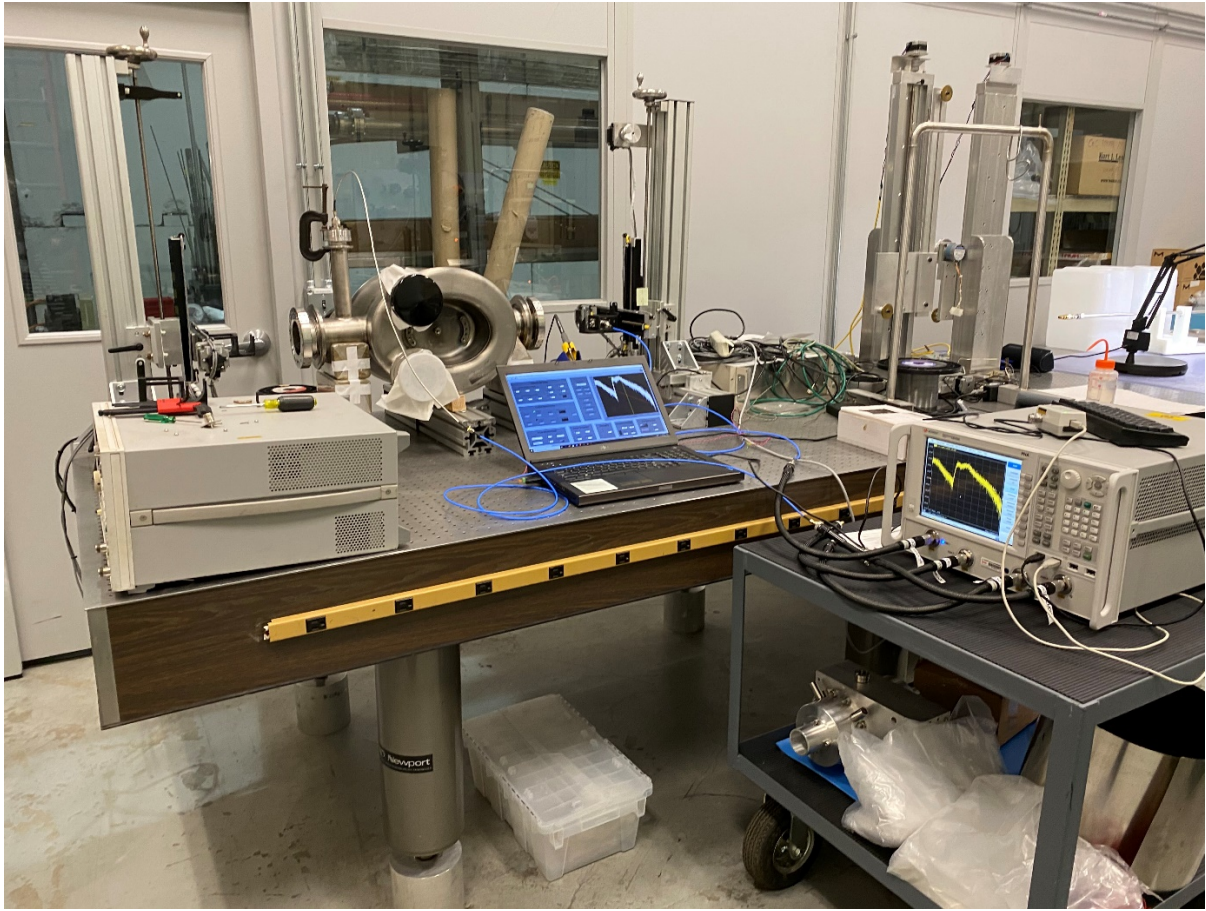

Figure 1: Stretched wire pulled through the DQW crab cavity beam pipe.

A copper hook-type antenna with $Q_{ext}$ in the order of $10^6$ was inserted into the port opening on the cavity beam pipe. This coupler antenna is used to pick up the RF signal excited in the cavity.

Two ends of the wire are fixed to an independent 2-way stage individually. The motor driven the stage has a step resolution of 0.01 mm. An automated LabView code was used to find the electrical center. The starting point of the wire location is manually measured geometric center of the beam pipe, and the stages moved in independently with a scan from coarse to fine grids in the area defined by the LabView program. The scan grid spaced by 0.5 mm, and data was taking at each grid point.

## RF Data with analysis:

The DQW crab cavity has unique field pattern at fundamental crabbing mode. The electrical field distributed evenly between the two deflecting plates.

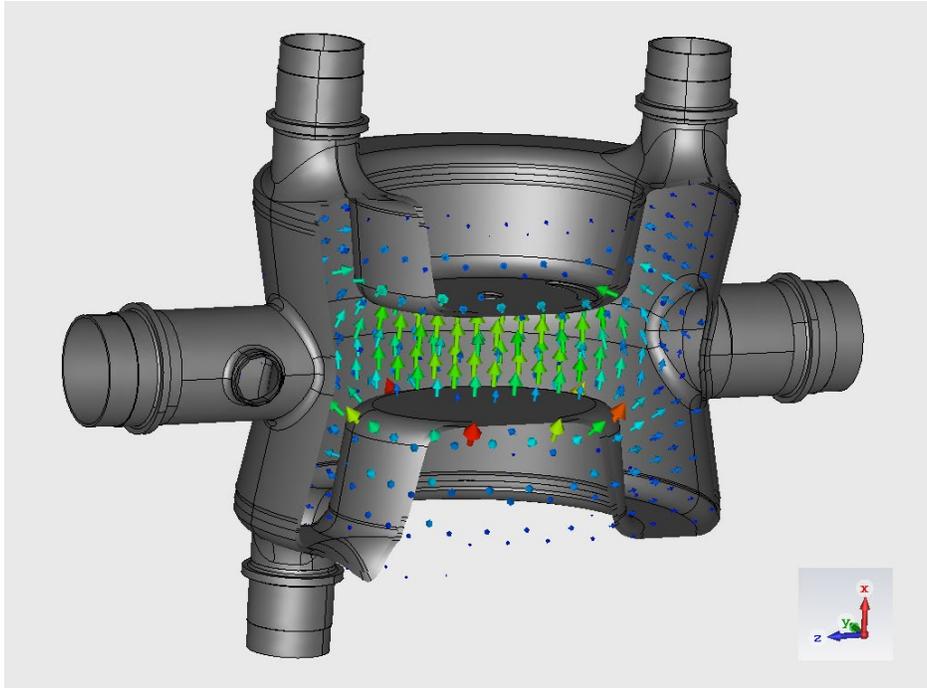

Figure 2: Electric field inside the DQW crab cavity at fundamental mode frequency of 400 MHz.

The initial location of the wire was changed manually to four different settings, which were picked by moving the wire along the symmetry plane. The crab cavity has high symmetry along the center deflecting plates, however the coupler ports broke the symmetry due to their different locations. The fabrication also introduced additional asymmetry which is small but random. The purpose of the measurement is to find the actual electrical center and use the result as a validation to the stretched wire system to these exotic type resonators.

Due to the symmetry of the EM field of the deflecting mode, the RF signal sent through the wire does not excited any resonance in the cavity when the wire runs along the cavity electrical center (EC) plane. As the wire offsets from the EC plane, the EM field builds up in the cavity and can be picked up by the copper antenna.

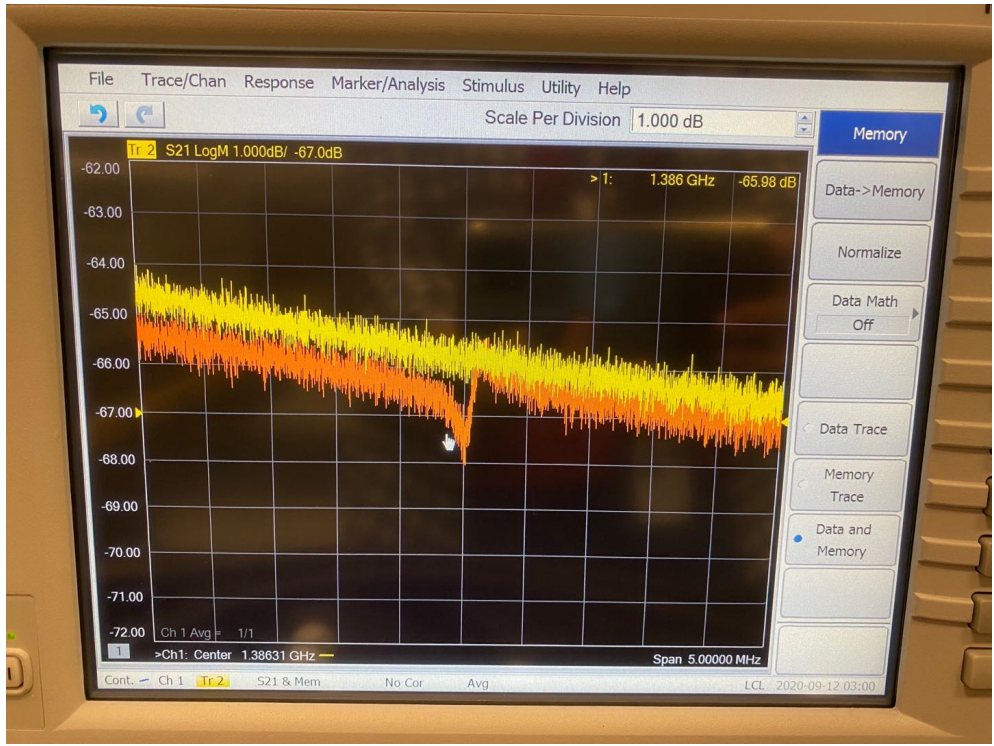

Figure 3: An example of signal scan from stretched wire at different locations: on E center plane (yellow), and off E center plane (orange)

The fine scanning program compares the peak of the signal with the baseline at each grid point, with the slope in the base being corrected. From each initial location, the wire scanned through an area of 10mm by 10mm surrounding and stopped automatically when the signal peak reached a local minimum. The length of the wire is 91.2cm.

The coordinates of four electrical centers are listed below:

Table 1: Electrical Center locations set by stretched wire scanning. Coordinates are relative to a preset origin of the motor; unit is in centimeter.

|      | X_stage1 | Y_stage1 | X_stage2 | Y_stage2 |
|------|----------|----------|----------|----------|
| EC 1 | 1.81737  | 38       | -134.25  | -30.22   |
| EC 2 | -22.95   | 37.33    | -130.13  | -30.22   |
| EC 3 | -31.317  | 36.9875  | -123.473 | -30.4851 |
| EC 4 | -34.2    | 37.3     | -139.6   | -30.5    |

The four locations of the wire were measured with respected to the cavity flange with professional survey FARO Arm.

## Survey:

Survey of the wire locations at the EC was completed by the Survey Group at Collider Accelerator Department of BNL based on the experience in machine alignment for RHIC and NSLS-2. At each verified EC location, both stages were scanned with FARO Arm laser scanner with a 1cm long wire from the stage surface.

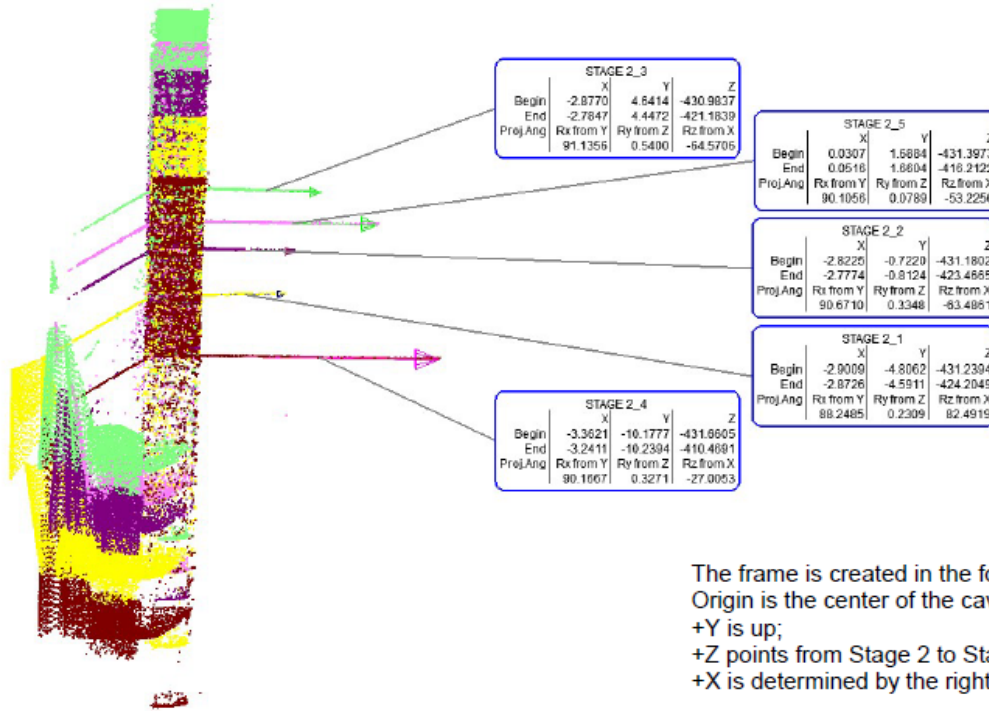

Figure 4: FIve laser scanned overlapping images reproduced in the FARO Arm data analysis software.

Figure 4 shows image scan of same side stage with 1cm wire extended at four different EC locations and one manually measured geometric center location. The center of the wire from the stage vertical surface are connected with the center point at the other side of the stage. The straight line defined by the two points are used to substitute the actual metal wire, and used to define the line in the EC plane.

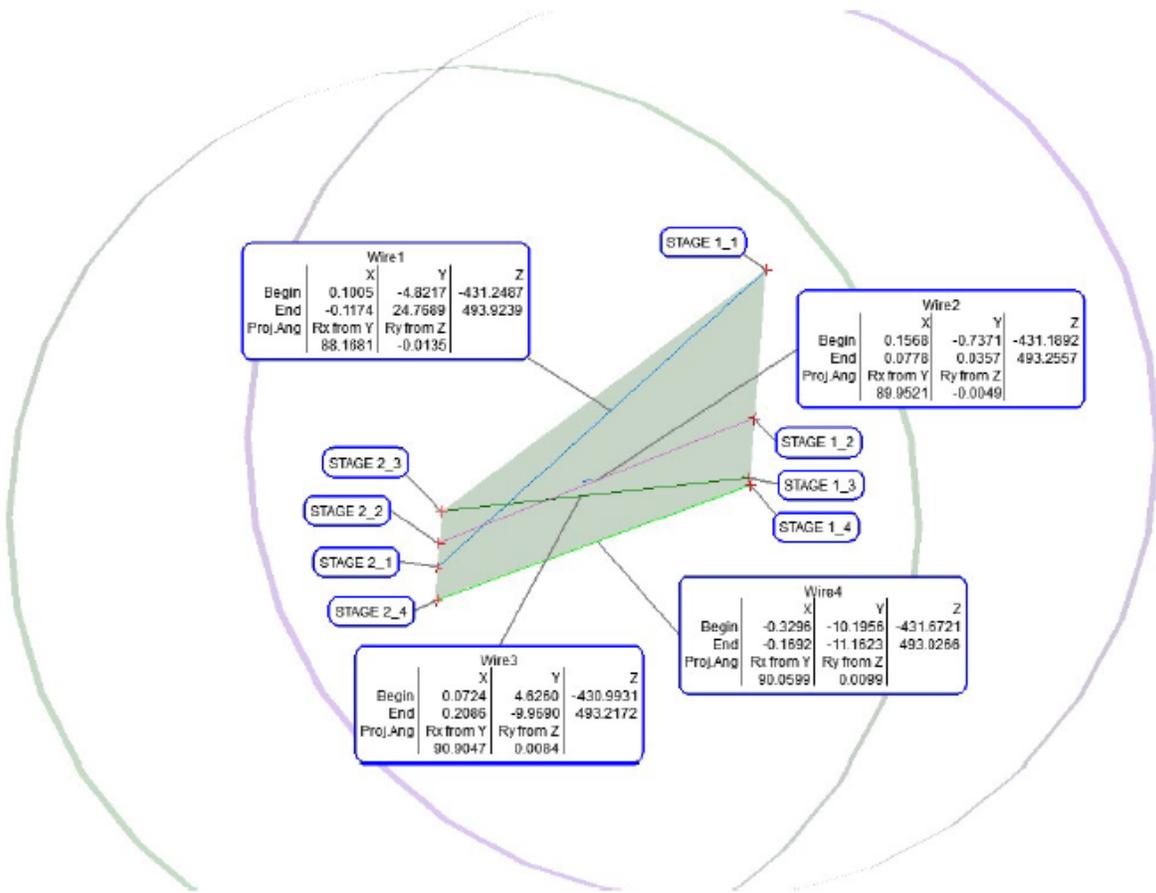

Figure 5: Four lines defined by the stretched wire measurements

The four lines defined by the four pairs of center points from stage 1 and stage 2 are marked in Figure 5. The plane fitted by the four lines is shown in green in the same figure. This is the EC plane of the cavity.

The survey showed that the EC plane of the cavity is along the symmetry plane of the cavity deflecting plates with small rotating angles listed in Table 2.

Table 2: Rotation angles between measured EC center plane and cavity symmetry plane. The axis are labeled as in Figure 2.

| Rotation angle along X axis (deg) | Rotation angle along Y axis (deg) | Rotation angle along Z axis (deg) |
|---|---|---|
| -0.0017 | -0.3143 | 0.3067 |

The error of the four EC wires with respect to the fitted EC plane is less than 2e-4 mm in relative distance at the stage, and less than 0.2mrad in rotation.

## Summary

The stretched wire system built at BNL has successfully measured four different electrical center trajectories in the EC plane of the crab cavity. The survey has sufficient resolution to capture the relative electrical center plane with the cavity surface. This validated the stretched wire measurement system at BNL, and provided experience for cavity alignment during installation phase.

## Reference

[1] H. Wang, *Wire Stretching Technique for Measuring RF Crabbing/Deflecting Cavity Electrical Center and a Demonstration Experiment on Its Accuracy*, Proceedings of NAPAC2016, Chicago, IL, USA, p 225-229.

[2] T. Xin, et.al., *Measuring the Electrical Center and Field Flatness of 704 MHz Deflecting Cavity for LEReC with Wire Stretching System,* Proceedings of IPAC2018, Vancouver, BC, Canada, p 1320-1322.